# Dynamics Investigation of the quantum-control-assisted multipartite uncertainty relation in Heisenberg model with Dzyaloshinski-Moriya interaction

Jie Xu, Xiao Zheng [*], Ai-Ling Ji, Guo-Feng Zhang [*]

*School of Physics, Beihang University, Beijing 100191, China*

**Abstract:** Recently, Zheng constructs a quantum-control-assisted multipartite variance-based uncertainty relation, which successfully extends the conditional uncertainty relation to the multipartite case [Annalen der physik, 533, 2100014 (2021)]. We here investigate the dynamics of the new uncertainty relation in the Heisenberg system with the Dzyaloshinski-Moriya interaction. It is found that, different from entanglement, the mixedness of the system has an interesting single-valued relationship with the tightness and lower bound of the uncertainty relation. This single-valued relationship indicates that the tightness and lower bound of the uncertainty relation can be written as the functional form of the mixedness. Moreover, the single-valued relationship with the mixedness is the common nature of conditional uncertainty relations, and has no relationship with the form of the uncertainty relations. Also, the comparison between the new conditional variance-based uncertainty relation and the existing entropic one has been made.

## I. Introduction

Uncertainty relation, a characteristic of the quantum system, is very useful for quantum information science [1-3]. Applying two incompatible variables to the uncertainty relation, the uncertainty of corresponding measurement results will be constrained by the lower bound, which provides a theoretical basis in quantum information science [4-11]. Many uncertainty relations have been constructed for smaller uncertainty, and these uncertainty relations generally can be divided into two types: uncertainty relations based on variance [12-17] and entropic uncertainty relations [18-19].

The most famous variance-based uncertainty relation is the Schrödinger uncertainty relation (SUR), which reads [3]:

$$\Delta A^2 \Delta B^2 \geq \frac{1}{4}|\langle[A,B]\rangle|^2 + \frac{1}{4}|\langle\{\check{A},\check{B}\}\rangle|^2, \qquad (1)$$

where $\Delta A^2$ ($\Delta B^2$) represents the variance of the observable $A$ ($B$), $[A,B] = AB - BA$, and $\{\breve{A},\breve{B}\} = \breve{A}\breve{B} + \breve{B}\breve{A}$ with $\breve{A} = A - \langle A \rangle$ ($\breve{B} = B - \langle B \rangle$) and $\langle A \rangle$ ($\langle B \rangle$) being the expected value of $A$ ($B$). The well-known entropic uncertainty relation is constructed by Kraus [18]:

$$H(R) + H(S) \geq \log_2 \frac{1}{c}, \tag{2}$$

where $R$ and $S$ are two incompatible observables, $H(R)$ ($H(S)$) represents the Shannon entropy of the observable $R$ ($S$), and $c = \max_{r,s} |\langle \varphi_r | \phi_s \rangle|^2$ with $|\varphi_r\rangle$ ($|\phi_s\rangle$) being the eigenvector of $R$ ($S$).

In 2010, Berta et al. constructed a quantum memory-assisted entropic uncertainty relation (QM-EUR) [19]:

$$H(R|B) + H(S|B) \geq \log_2 \frac{1}{c} + H(A|B), \tag{3}$$

where $A$ is the measured system, $B$ represents the memory system. $H(A|B) = H(\rho_{AB}) - H(\rho_B)$ stands for the conditional von Neumann entropy of the density operator $\rho_{AB}$ with $\rho_{AB}$ being the state of the whole system and $H(\rho)$ being Shannon entropy of the state $\rho$. $H(Y|B) = H(\rho_{YB}) - H(\rho_B)$ represents the conditional von Neumann entropy of the density operator $\rho_{YB}$ with $\rho_{YB} = \sum_Y (|\varphi_Y\rangle\langle\varphi_Y| \otimes I) \rho_{AB} (|\varphi_Y\rangle\langle\varphi_Y| \otimes I)$ being the post-measurement state. $|\varphi_Y\rangle$ is the eigenvector of $Y$ with $Y \in (R,S)$ and $I$ stands for identical operator. $\rho_B = Tr_A(\rho_{AB})$ is the density operator of the memory system. $H(R|B)$ ($H(S|B)$) represents the uncertainty of the measurement $R$ ($S$) conditioned on the prior information stored in memory system $B$.

QM-EUR indicates that the traditional entropy uncertainty relation (2) can be broken with the help of a quantum memory system. In fact, QM-EUR is essentially a conditional uncertainty relation based on entropy [19], and its construction opens up a new direction for the understanding of uncertainty relation [20-22]. However, due to the complexity of conditional entropy, the investigation of the conditional uncertainty relation develops slowly [23].

To fix this problem, Zheng recently constructed a conditional uncertainty relation in form of variance, which is named as quantum control-assisted variance uncertainty relation (QC-VUR), and showed that the traditional variance uncertainty relation can be broken with the quantum control systems [23]. The obtained uncertainty relation can be illustrated by the following game: 1. Bob prepares $N+1$ entangled particles, which are denoted by $A, C_1, C_2, \cdots, C_N$. $A$ is called measured system and $C_1, C_2, \cdots, C_N$ are control systems; 2. Bob sends measured system $A$ to Alice; 3. Alice selects a measurement to be taken, such as $Q_k$, and tells it to Bob; 4. Relying on the information

Bob and Alice have about the quantum state of the whole system, Bob chooses an appropriate measurement, such as $O_k$, and implements $O_k$ on control systems $C_1, C_2, \cdots, C_N$, respectively; 5. Alice implements $Q_k$ on the measured system $A$.

The measurements performed on control systems $C_1, C_2, \cdots, C_N$ can be used to control the whole system for the smallest uncertainty of the measured system $A$. Therefore, the measurements implemented on the corresponding control systems are called quantum control. QC-VUR is essentially a conditional uncertainty relation, and can be expressed by the following inequality [23]:

$$\sum_{k=1}^{K} E[V(Q_k^A | O_k^{C_1}, \cdots, O_k^{C_N})] \geq L_{tra} - \sum_{k=1}^{K} V[E(Q_k^A | Q_k^{C_1})]$$
$$- \sum_{k=1}^{K} \sum_{n=2}^{N} E[V(E[Q_k^A | O_k^{C_n}] | O_k^{C_1}, \cdots, O_k^{C_{n-1}})], \quad (4)$$

where $O_k^{C_n}$ ($Q_k^A$) represents the measurement $O_k$ ($Q_k$) performed on particle $C_n$ ($A$). $E[V(Q_k^A | O_k^{C_1}, \cdots, O_k^{C_n})]$ is the conditional variance of $Q_k^A$ under the condition that the measurements $O_k^{C_1}, \cdots, O_k^{C_n}$ have been performed. $V[E(Q_k^A | O_k^{C_1})]$ is defined as the variance of $E(Q_k^A)$ conditioned that one has performed the measurement $O_k^{C_1}$, and $E[V(E(Q_k^A | O_k^{C_n})] | O_k^{C_1}, \cdots, O_k^{C_{n-1}})]$ represents the conditional variance of $E(Q_k^A | O_k^{C_n})$ on the condition that the measurements $O_k^{C_1}, \cdots, O_k^{C_{n-1}}$ have been performed [23]. The construction of the QC-VUR successfully extends the conditional uncertainty relation to the multipartite case, and promotes the development of uncertainty relation.

In Ref. [24], Zheng et al. investigated QM-EUR in the Heisenberg system with Dzyaloshinski-Moriya (DM) interaction, and concluded that, compared with entanglement, the mixedness has a closer relationship with QM-EUR. Moreover, the tightness of QM-EUR can be written as the function of the mixedness, which brings us a new understanding of the QM-EUR and the mixedness. As mentioned above, both QM-EUR and QC-VUR are essentially conditional uncertainty relations, and a natural question then raises that does there exist the similar relationship between QC-VUR and the mixedness.

To this aim, in this paper we mainly investigate the influence of the mixedness and the entanglement on the QC-VUR in the Heisenberg system with DM interaction. Also, the comparison between the QM-EUR and QC-VUR is made. The remainder paper is divided into four sections. Sec. II is devoted to the introduction of the physical model of the Heisenberg system with DM interaction. In Sec. III, the dynamics properties of QC-VUR and relationship between QC-VUR and the mixedness are investigated. The comparison between QC-VUR and QM-EUR is made in Sec. IV. Finally, Sec. V is devoted to the conclusion.

## II. Heisenberg Model with DM Interaction

The Hamiltonian of Heisenberg model with DM interaction can be expressed:

$$H_{DM} = \frac{J}{2}[(\sigma_{1x}\sigma_{2x} + \sigma_{1y}\sigma_{2y} + \sigma_{1z}\sigma_{2z}) + D \cdot (\sigma_{1x}\sigma_{2y} - \sigma_{1y}\sigma_{2x})]$$

$$= J[(1+iD)\sigma_{1+}\sigma_{2-} + (1-iD)\sigma_{1-}\sigma_{2+}], \quad (5)$$

where $J$ is the real coupling coefficient, the state is a antiferromagnetic one when $J > 0$ and $J < 0$ for the ferromagnetic case. $\vec{D} = D\vec{z}$ is the DM coupling vector [25-26]. The quantum state of the system becomes $\rho(T) = e^{-\beta H}/Z$ when the system reaches the thermal equilibrium state, where $H$ is the Hamiltonian of the system, $T$ represents the temperature of the system, and $Z = Tr(e^{-\beta H})$ is the partition function. Based on the discussion above, in thermal equilibrium state, the density matrix $\rho(T)$ of the Heisenberg model with DM interaction can be obtained [24]:

$$\rho(T) = \frac{1}{Z}\begin{pmatrix} \rho_{11} & 0 & 0 & 0 \\ 0 & \rho_{22} & \rho_{23} & 0 \\ 0 & \rho_{23}^* & \rho_{33} & 0 \\ 0 & 0 & 0 & \rho_{44} \end{pmatrix}, \quad (6)$$

with the elements $\rho_{11} = \rho_{44} = e^{-\beta J/2}$, $\rho_{22} = \rho_{33} = e^{\beta(J-\delta)/2}(1+e^{\beta\delta})/2$, $\rho_{23} = e^{i\theta}e^{\beta(J-\delta)/2}(1-e^{\beta\delta})/2$, $\rho_{23}^* = e^{-i\theta}e^{\beta(J-\delta)/2}(1-e^{\beta\delta})/2$, $Z = 2e^{-\beta J/2}(1+e^{\beta J}\cosh(\beta\delta/2))$, $\beta = 1/kT$, and $\delta = 2J\sqrt{1+D^2}$.

## III. Effects of Mixedness and Entanglement on the QC-VUR

In this section, the effects of the entanglement and the mixedness on QC-VUR will be investigated. For simplification, Boltzmann constant is taken as $k = 1$. At finite temperature, the entanglement of the two-qubits Heisenberg system with DM interaction can be measured by concurrence $C$, which reads [27]:

$$C = \frac{2}{Z}\max\left[\frac{1}{2}\left|e^{\frac{\beta(J-\delta)}{2}}(1-e^{\beta\delta})\right| - e^{\frac{\beta J}{2}}, 0\right]. \quad (7)$$

As we know, the quantity $Tr(\rho^2)$ can be used to distinguish the state of $\rho$. i.e., $Tr(\rho^2) = 1$ indicates that the state is pure one, and $Tr(\rho^2) < 1$ for the mixed one. Denoting $1 - Tr(\rho^2)$ by $\gamma$, it can be deduced that the greater $\gamma$ is, the stronger the mixedness of $\rho$ is. Thus, $\gamma$ can be employed to measure the mixedness of the system [28-29]. According to the definition of the mixedness and Eq. (6), the mixedness of Heisenberg system with DM interaction is obtained [24]:

$$\gamma = \frac{4e^{\beta(J+\delta)}[\cosh(\beta J) + 2\cosh\left(\frac{\beta\delta}{2}\right)]}{[e^{\beta(J+\delta)} + e^{\beta J} + 2e^{\frac{\beta\delta}{2}}]^2}. \quad (8)$$

The traditional lower bound $L_{tra}$ in QC-VUR is taken as the lower bound of the uncertainty relation constructed in Ref. [30]:

$$\Delta A^2 + \Delta B^2 \geq L_{tra} = \frac{|\langle O^\dagger(\check{A}+e^{i\theta}\check{B})\rangle|^2}{\langle O^\dagger O\rangle} - \langle\{\check{A}, e^{i\theta}\check{B}\}_G\rangle, \tag{9}$$

where $\{\check{A}, e^{i\theta}\check{B}\}_G = \check{A}^\dagger e^{i\theta}\check{B} + e^{-i\theta}\check{B}^\dagger\check{A}$, $\theta \in [0,2\pi]$, and $O$ represents an arbitrary operator.

In order to investigate the properties of QC-VUR in the Heisenberg model, we define [24]:

$$W = L_{tra} - \sum_{k=1}^{K} V\left[E\left(Q_k^A|O_k^{C_1}\right)\right] - \sum_{k=1}^{K}\sum_{n=2}^{N} E[V(E[Q_k^A|O_k^{C_n}]|O_k^{C_1},\cdots,O_k^{C_{n-1}})]. \tag{10}$$

$$U = \sum_{k=1}^{K} E[V(Q_k^A|O_k^{C_1},\cdots,O_k^{C_N})]/W. \tag{11}$$

$W$, the lower bound of QC-VUR, is used to measure the quality of the uncertainty relation, i.e., the smaller the value of $W$ is, the better the quality of the uncertainty relation is. $W = 0$ indicates that all $Q_k^A$ can be accurately predicted at the same time [31]. $U$ is the ratio of the left to right side of the uncertainty relation, which can be employed to measure the tightness of the uncertainty relation. In the following, the dynamics of the uncertainty relation will be investigated from the perspective of $W$ and $U$, respectively.

Firstly, the investigation of the QC-VUR is focused on the lower bound $W$. In the Heisenberg model with DM interaction, the evolution of $W$, $C$ and $\gamma$ with respect to $D$ and $J$ is shown in Fig.1. It can be seen from Fig.1 (a), (b), (e) and (f) that the stronger the entanglement of the system is, the smaller the lower bound of the uncertainty relation is, which indicates that the entanglement between the control systems and the measured system can effectively reduce the uncertainty of the measurement results. Also, from Fig.1 (c), (d), (e) and (f), one can see that the stronger the mixedness of the system is, the greater the lower bound of the uncertainty relation is. That is to say, different from entanglement, the mixedness of the system has a positive relationship with the lower bound.

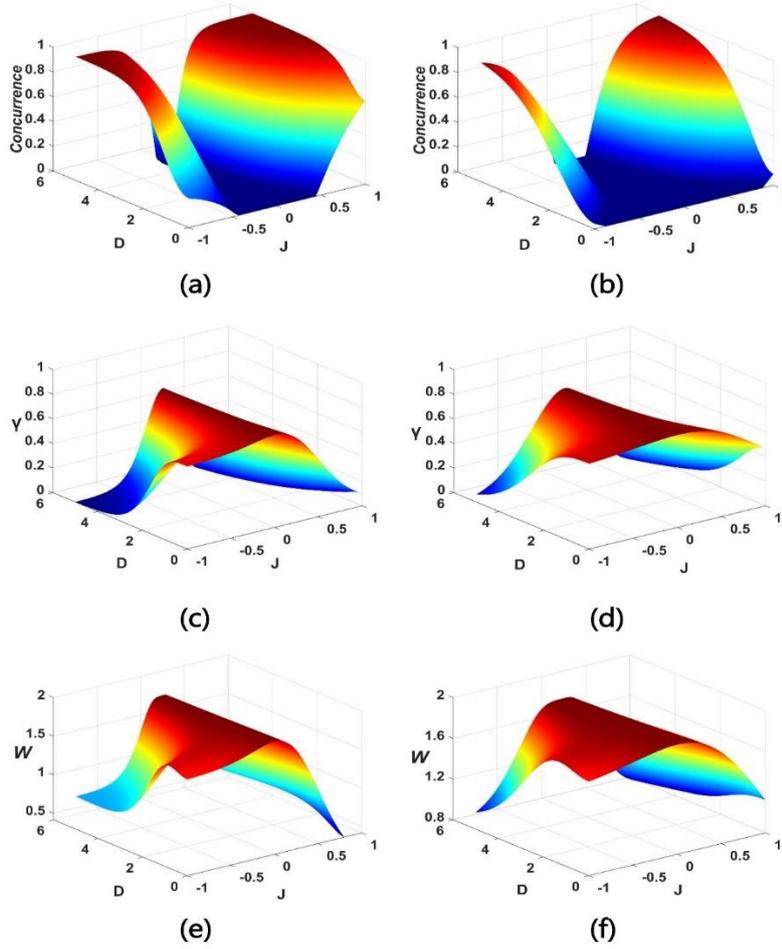

Fig.1 Evolution of $C$, $\gamma$ and $W$ with respect to $D$ and $J$ for $T = 0.5$ in (a), (c) and (e), for $T = 1$ in (b), (d) and (f). Here $K = 2$, $Q_1 = O_1 = \sigma_x$, $Q_2 = O_2 = \sigma_z$, $O = \sigma_x + \sigma_z$, and $\theta = 0.5$.

In addition, one can see from Fig.1 (c), (d), (e) and (f) that the evolution of $W$ is similar to that of $\gamma$, which means that, compared with entanglement, the mixedness has a closer relationship with the lower bound. The further investigation of this relationship between the lower bound and mixedness is made in Fig.2. From Fig.2, one can see that the lower bound and mixedness share a similar structure of evolution, and there exists an obvious single-valued function relationship between the lower bound and the mixedness. Therefore, one can obtain that the mixedness has a closer relationship with the lower bound than the entanglement.

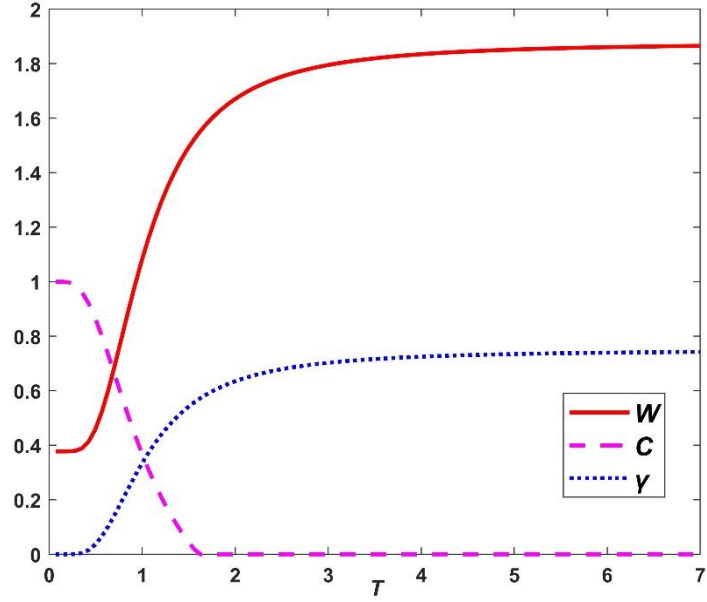

Fig.2 Evolution of $W, C, \gamma$ with $T$, and we here take $D = 1, J = 1, O = \sigma_x + \sigma_z$ and $\theta = 0.5$.

As we know, the single-valued relationship between the lower bound $W$ and the mixedness $\gamma$ indicates that the lower bound can be written as a function of the mixedness. After some simple verification, we find that the lower bound $W$ is actually a function with respect to $(D, \gamma)$. It should be mentioned that the corresponding functional form depends only on the sign of $J$, and has nothing to do with the specific value of $J$ and $T$, as shown in Fig.3. Also, from Fig.3 (a), we can see that the lower bound $W$ vanishes when the $\gamma$ and $D$ are both equal to zero. That is to say, in the Heisenberg model without the DM interaction, the measurement results of the pure state, i.e. the minimum-mixedness state, can be accurately predicted in antiferromagnetic case.

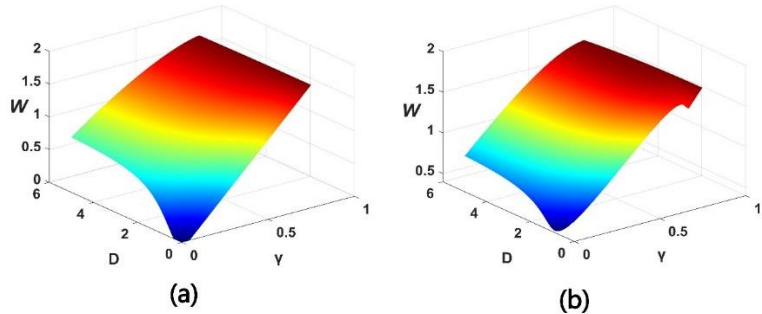

Fig.3 Evolution of $W$ with $(\gamma, D)$ for $J = 1$ in (a) and $J = -1$ in (b)
$O = \sigma_x + \sigma_z, \theta = 0.5$.

Then, the QC-VUR is investigated from the perspective of the tightness $U$. The evolution of $U$ with respect to $D$ and $J$ is shown in Fig.4.

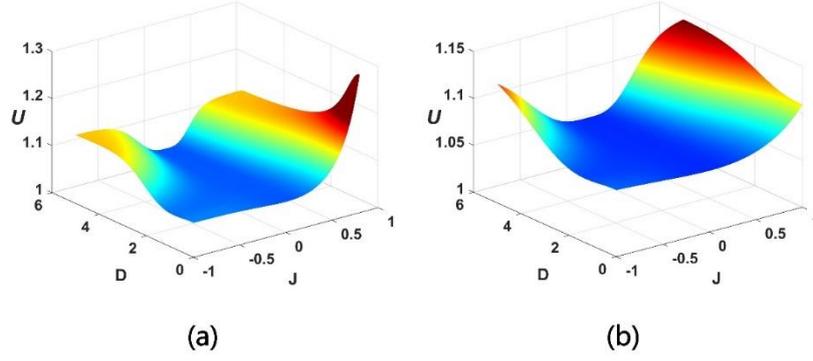

Fig.4 Evolution of $U$ with $(D, J)$ for $T = 0.5$ in (a) and $T = 1$ in (b)

$O = \sigma_x + \sigma_z, \theta = 0.5$.

Comparing Fig.1 (a), (b) with Fig.4 (a), (b), one can see that the stronger the entanglement is, the worse the tightness is. That is to say, the entanglement between the control systems and the measured system can destroy the tightness of the uncertainty relation. Also, by comparing Fig.1 (c), (d) with Fig.4 (a), (b), it can be seen that the greater $\gamma$ is, the smaller $U$ is, which means that the tightness of system will be improved by the increase of the mixedness. And the evolution of $U$ has an opposite structure with that of $\gamma$. That is to say, compared with entanglement, the mixedness has a closer relationship with the tightness. Further investigation of the relationship between the tightness and mixedness is made in Fig.5.

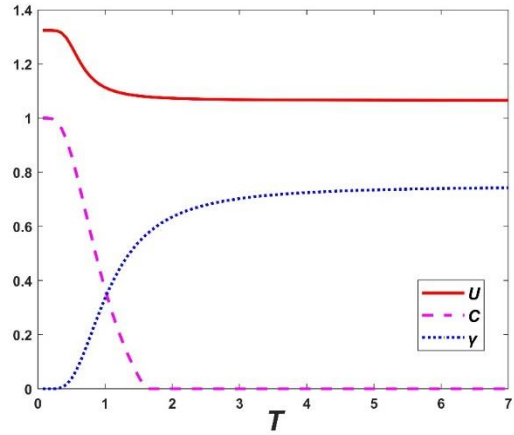

Fig.5 Evolution of $U, C, \gamma$ with $T$, and we here take $D = 1, J = 1$

$O = \sigma_x + \sigma_z, \theta = 0.5$.

In Fig.5, there is an obvious single-valued function relationship between the tightness and the mixedness. Therefore, we hold the view that the mixedness has a closer relationship with the tightness than the entanglement.

The evolution of tightness with respect to $\gamma$ and $D$ is shown in Fig.6. One can clearly see that $U$

and $\gamma$ have a single-valued function relationship, which confirms the close relationship between the tightness and the mixedness.

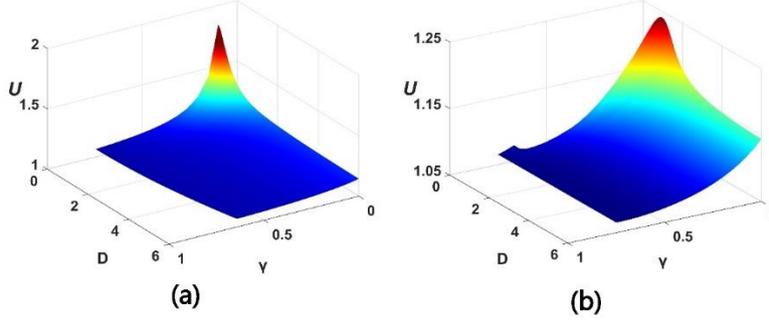

Fig.6 Evolution of $U$ with $(\gamma, D)$ for $J = 1$ in (a) and $J = -1$ in (b)
$O = \sigma_x + \sigma_z, \theta = 0.5$.

Notably, according to Ref. [24], one can find that the similar conclusion that the uncertainty relation has single-valued relationship with mixedness is also obtained for QM-EUR. That is to say, both QM-EUR and QC-VUR have single-valued relationship with mixedness. Considering that the two uncertainty relations are both conditional uncertainty relations, one can in fact obtain that the single-valued relationship between the mixedness and conditional uncertainty relation does not rely on the form of the uncertainty relation

## IV. Comparison between QC-VUR and QM-EUR

Based on the discussion above, one can see that both QC-VUR and QM-EUR are conditional uncertainty relations, which indicates that the traditional uncertainty relation can be broken with the help of entanglement. The difference between them is that QC-VUR is based on quantum control in the form of variance and QM-EUR is based on quantum memory from the perspective of entropy. Then a natural question raises that which of them can express the essence of conditional uncertainty relation better. Thus, a comparison between QC-VUR and QM-EUR is made in this section.

Tightness, the ratio of the left to right side of uncertainty relation, is a relatively reasonable quantity to quantify the performance of different uncertainty relations types, and we thus define the tightness of the QM-EUR as [24]:

$$U = [H(R|B) + H(S|B)]/[\log_2 \frac{1}{c} + H(A|B)]. \tag{12}$$

Using Eq. (11) and Eq. (12), one can obtain the evolutions of QC-VUR and QM-EUR with respect to $J$ and $D$, as shown in Fig.7. From Fig.7, it can be seen that the tightness of QC-VUR is better than that of QM-EUR in most area, and we thus obtain that the tightness of QC-VUR is better than that of QM-EUR in the Heisenberg Model with DM Interaction.

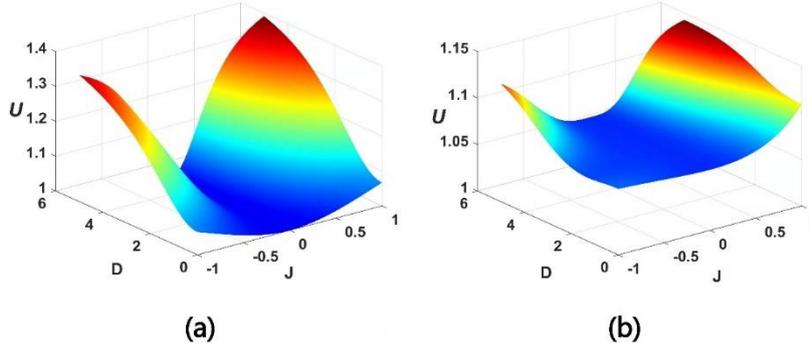

Fig.7 Evolution of $U$ with $D, J$ for QM-EUR in (a) and QC-VUR in (b), and here $T = 1$
$O = \sigma_x + \sigma_z, \theta = 0.5$.

## V. Conclusions

To conclude, we have investigated the dynamics of QC-VUR in Heisenberg system with DM interaction, and made a comparison between QC-VUR and QM-EUR. Firstly, the uncertainty of measured system will be reduced by the entanglement between measured system and control systems, while be improved by the mixedness of the whole system. Moreover, the measurement results can be accurately predicted when the system is in the minimum-mixedness state. In contrast to the uncertainty, the tightness will be destroyed by the entanglement, whereas the mixedness can make the uncertainty relation tighter. Further investigation found that, compared with entanglement, mixedness has a closer relationship with the conditional uncertainty relation. There exists a single-valued relationship between mixedness and the conditional uncertainty relation, which means that the lower bound and the tightness of the conditional uncertainty relation can be written as a function of the mixedness, and the single-valued relationship has no relation with the form of conditional uncertainty relation. Finally, in the comparison of the tightness, QC-VUR performs better than QM-EUR in this model.

**Acknowledgments**

This work is supported by the National Natural Science Foundation of China (Grant No. 12074027).

**Data Availability Statements**

Data sharing not applicable to this article as no datasets were generated or analysed during the current study.